%% file: main.tex
\documentclass[a4paper,fleqn]{cas-dc}

\usepackage[numbers]{natbib}
\usepackage{paralist}

\def\tsc#1{\csdef{#1}{\textsc{\lowercase{#1}}\xspace}}
\tsc{WGM}
\tsc{QE}
\tsc{EP}
\tsc{PMS}
\tsc{BEC}
\tsc{DE}

\begin{document}
\let\WriteBookmarks\relax
\def\floatpagepagefraction{1}
\def\textpagefraction{.001}
\shorttitle{Assets in Software Engineering: What are they after all?}
\shortauthors{E. Zabardast et~al.}

\title [mode = title]{Assets in Software Engineering: What are they after all?}                      

\tnotetext[1]{This research was supported by the KK foundation through the SHADE KK-H\"{o}g project (Ref. 2017/0176) and Research Profile project SERT (Ref. 2018/010)  at Blekinge Institute of Technology, SERL Sweden.}


\author[1]{Ehsan Zabardast}[orcid=0000-0002-1729-5154]
\cormark[1]
\ead{ehsan.zabardast@bth.se}
\ead[url]{https://www.ehsanzabardast.com}

\address[1]{Software Engineering Research Lab SERL,
            Blekinge Institute of Technology,
            Campus Karlskrona, Valhallav{\"a}gen 1, 
            Karlskrona, Sweden }

\author[1]{Julian Frattini}[orcid=0000-0003-3995-6125]
\ead{julian.frattini@bth.se}

\author[1]{Javier Gonzalez-Huerta}[orcid=0000-0003-1350-7030]
\ead{javier.gonzalez.huerta@bth.se}
\ead[URL]{http://www.gonzalez-huerta.net}

\author[1,2]{Daniel Mendez}[orcid=0000-0003-0619-6027]
\ead{daniel.mendez@bth.se}
\ead[URL]{http://www.mendezfe.org/}
\address[2]{fortiss GmbH, Guerickestraße 25, 80805 Munich, Germany}

\author[1,2]{Tony Gorschek}
\ead{tony.gorschek@bth.se}
\ead[URL]{http://gorschek.com/}

\author[1]{Krzysztof Wnuk}
\ead{krzysztof.wnuk@bth.se}

\cortext[cor1]{Corresponding author}

\begin{abstract}
During the development and maintenance of software-intensive products or services, we depend on various artefacts. Some of those artefacts, we deem central to the feasibility of a project and the product's final quality. Typically, these central artefacts are referred to as assets. However, despite their central role in the software development process, little thought is yet invested into what eventually characterises as an asset, often resulting in many terms and underlying concepts being mixed and used inconsistently. A precise terminology of assets and related concepts, such as asset degradation, are crucial for setting up a new generation of cost-effective software engineering practices.

\noindent In this position paper, we critically reflect upon the notion of \textit{assets in software engineering}. As a starting point, we define the terminology and concepts of assets and extend the reasoning behind them. We explore assets' characteristics and discuss what \textit{asset degradation} is as well as its various types and the implications that asset degradation might bring for the planning, realisation, and evolution of software-intensive products and services over time.  

\noindent We aspire to contribute to a more standardised definition of \textit{assets in software engineering} and foster research endeavours and their practical dissemination in a common, more unified direction.
\end{abstract}




\maketitle

\input{body}




\bibliographystyle{cas-model2-names}

\bibliography{references}




\end{document}

%% file: body.tex
\section{Introduction}

A fundamental challenge in producing software-intensive products and services is coping with the continuous changes that the business environments and the customers demand from the products~\cite{casale2016current}. These frequent changes have consequences on software artefacts, the main building blocks of the development process of software-intensive products and services~\cite{MendezFernandez2019}. 

A software artefact is defined as ``a work product that is produced, modified, or used by a sequence of tasks that have value to a role~\cite{MendezFernandez2019}.'' This is a broad definition, and the number of artefacts produced or used along the development and maintenance activities of software-intensive products and services may be extensive. However, in our perception, software artefacts related to source code and architecture seem to be more prominent in literature and better understood than other artefacts like manuals or requirements, despite the latter having an at least comparable impact on the development life cycle.

Although all software artefacts are subject to quality degradation, it is neither practical nor efficient to exercise control of the quality for all of them. Quality control is strengthened by continuous use. In our perception, continuously controlling and ensuring quality is justified if an artefact is \textit{intended to be used more than once} over time, i.e., if it is used several times during the development or maintenance activities. This, as we will argue, is one of the key characteristics of the related term \textit{``asset''}. However, software engineering lacks a profound understanding of the term and its implications for software engineering practices.

Assets and asset management are popular terms used outside of software engineering. Reviewing the literature reveals that, despite the popularity of the term (see, e.g.,~\cite{joe2010,schneider2006asset}), often related to tangible (physical) systems, it has not yet received much attention in software engineering. Nonetheless, we argue for the importance of asset management in software engineering as well as its evolution, and we postulate the importance of a well-defined vocabulary.

In this position paper, we critically discuss and characterise assets and extend the reasoning around related concepts such as asset degradation, i.e., the loss of value of an asset. We discuss various types of value degradation and the possible implications of these on the planning, realisation, and evolution of software-intensive products and services. Our contribution is rooted in our long-term academia-industry collaborations\footnote{See, for instance, \url{www.rethought.se}.}, and the empirical studies we conducted together with our industrial partners. The novelty of this work is three-fold:
\begin{itemize}
    \item Characterising assets. Though there is a definition for assets (ISO 55000~\cite{ISO/IEC/IEEE2014}), there is no good way of distinguishing what an asset is and what it is not. Characterising assets can help identify them.
    \item Categorising the types of degradation. Discussing and defining different types of degradation can help practitioners and researchers better understand and investigate assets' degradation.
    \item Delineating from related concepts such as Technical Debt. As consequence of our continuous work and collaboration with the industry, we have observed that practitioners often  have a different perspective on Technical Debt (TD) than academic authors. Other assets that go beyond code-related artefacts are of importance for practitioners and the industry is concerned with the consequences of their continuous use (which indicates that it might be worth keeping control over). Highlighting this industrial perspective helps direct the research efforts to understand the needs of industry and validate our work.
\end{itemize}

Looking at related terms in software engineering literature, it seems that when referring to assets in software development, the term TD~\cite{Avgeriou2016,Cunningham1993} is becoming a \textit{catch-for-all} that works for all and every negative consequence that may or may not happen to assets. TD can be a metric that can help us measure the degradation of assets. Yet, TD is often mistaken for actual degradation due to the lack of a concise definition of asset degradation. Moreover, TD has not yet considered the propagation of degradation~\cite{Alegroth2017}, i.e. the degradation of certain assets stemming from the degradation of other, interrelated assets. Two examples of this propagation are i) the degradation of an architectural description element such as an activity diagram can lead to the degradation of source code, manifested as code-level TD; or ii) the propagation of code deprecation, which can propagate to the test-base, as in \cite{Sundelin2020}.

We conclude the paper with a discussion of future perspectives for research and practice to foster contemporary research endeavours in the community of software engineering researchers and practitioners in a common, more unified direction.

\section{Assets in Software-Intensive Products and Services}

In this section, we describe characteristics of assets, explore asset degradation and its propagation, and define asset management. We conclude this section by discussing our work in the context of software evolution.

\subsection{Assets' Characteristics}
\label{sec:characteristics}

An \textit{asset} is a software artefact that is intended to be used more than once during the inception, development, delivery, or evolution of a software-intensive product or service. Assets have ``potential or actual'' value to the organisation~\cite{ISO/IEC/IEEE2014}. Their value can be tangible or intangible, and financial or non-financial~\cite{ISO/IEC/IEEE2014}.

\begin{figure}[ht]
\centering
  \includegraphics[width=0.47\textwidth]{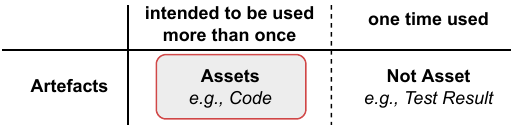}
\caption{Difference between \textit{assets} and \textit{not assets}. Assets are artefacts that are intended to be used more than once along the software development life cycle.}
\label{fig1:assets_v_artefacts}   
\end{figure}

If an artefact is used once and discarded or disregarded afterwards, it does not qualify as an asset. In our understanding, an asset is, per definition, an artefact, but not necessarily the other way around (see Figure~\ref{fig1:assets_v_artefacts}).

An artefact must be intended to be used more than once along the software development life cycle to be evolved and justify maintenance. However, this intention might not be apparent to the involved stakeholders; hence even an artefact that was originally assumed to be used only once but later actually used more than once qualifies as an asset. If an artefact has value for the organisation, the implication is that there is a need for controlling the artefact's quality. One-time-used artefacts, as we argue, do not have any relevance over time, rendering maintenance and evolution inefficient. In the paragraphs below, we provide a few examples:

\vspace{2mm}
\noindent\textbf{\textit{Artefacts that Qualify as Assets}}
\begin{itemize}
    \item \textit{Organisation Structure}: A document that summarises the roles involved in the development and their responsibilities in the organisation is an artefact that is intended to be used more than once because it is accessed whenever a specific competence needs to be located.
    \item \textit{Activity Diagram}: A visualisation of the system's behaviour is an artefact that is intended to be used more than once if it is used or accessed during the design and the development phases.
    \item \textit{Code}: Manually-written code is an artefact that is intended to be used more than once because it is used more than once. Each individual commit can be seen as a usage of the artefact.
\end{itemize} 

\noindent\textbf{\textit{Artefacts that Do Not Qualify as Assets}}
\begin{itemize}
    \item \textit{Generated View}: An automatically generated view (of parts) of the system that is only used once to capture the current state of the system is not an asset since it will be regenerated after the next system modification.
    \item \textit{Meeting Notes}: Meeting notes are one-time-used artefacts when they are used to create properly formatted meeting minutes and afterwards discarded.
    \item \textit{Test Result}: Test results can occur as one-time-used artefacts if they ---when deemed important--- are translated into a ticket or issue, while the test result artefact is discarded and not used again. 
\end{itemize}

It is important to highlight that this qualification as an asset is context-dependent. What counts as an asset for an organisation depends on, to name a few aspects, what product or service they deliver, the organisation's ways of working, and which artefacts are reused and which are not. For example: in the context of a company where UML model views are automatically extracted from the code and never modified, the code qualifies as an asset, but the extracted views do \textit{not} qualify as assets due to their singular use. On the contrary, consider a different company where the code is automatically generated from models through model transformations, and the code is never modified. In this second example, the code does \textit{not} qualify as an asset since it is only used once, but the models and the model transformations responsible for code generation qualify as assets in that particular organisation.

\subsection{Asset Degradation}

Software engineering practices commonly imply making trade-offs in solutions of competing qualities. The option to go for the highest possible quality might lead to over-engineering the solution. Is it always worth deciding for the highest quality solution, or is a ``good-enough'' solution preferable (whatever this eventually is and however this might eventually be measured)?

When referring to software development and assets, the term ``technical debt'' comes to mind. However, TD research mainly focuses on code-related assets~\cite{rios2018tertiary} and tends to depict the consequences of non-optimal design decisions~\cite{Avgeriou2016}. Moreover, TD does not consider the propagation of ``debt'' from one artefact to other artefacts~\cite{Alegroth2017}. Rios et~al.~\cite{rios2018tertiary} discussed the three main reasons why TD is mainly studied on code-related assets:

\begin{enumerate}[i]
    \item ``The concept of TD was initially coined by Ward Cunningham with a focus on coding activities [..] and this, in some way, may have influenced the initial directions of research in the area.~\cite{rios2018tertiary}''
    \item ``There is already a great amount of work that investigates the quality of software from indicators collected from the source code of the projects. Tools already available for this purpose may have been used as the starting point of the research community to analyse how debt can affect software projects.~\cite{rios2018tertiary}''
    \item ``The types of debt related to the code (architecture, design, code, defect, test) tend to cause effects that can be felt more quickly by the development team.~\cite{rios2018tertiary}''
\end{enumerate}

The degradation of certain assets, e.g., code, has been widely studied \cite{rios2018tertiary}. Such assets are more tangible and we can measure them to see how much they are deviating from the ---often purely academic--- gold standard.

We define degradation as the loss of value that an asset suffers due to intentional or unintentional decisions caused by technical or non-technical manipulation of the asset, or associated assets, during all stages of the product life-cycle. All assets can degrade, which will affect their \textit{value} ---for example, their usability~\cite{khurum2013software}--- in different ways.

We define three different types of degradation.\footnote{Fowler presents a similar classification for the debt metaphor at \url{https://martinfowler.com/bliki/TechnicalDebtQuadrant.html}} This classification was created as a result of discussions in a series of workshops with our industrial partner companies working on product development, including, for instance, Ericsson and Volvo CE, as part of a long-term academia-industry collaboration (\url{rethought.se}). We classify degradation as \textit{Deliberate}, \textit{Unintentional}, and \textit{Entropy}:

\begin{itemize}
    \item \textbf{Deliberate}: Taking a conscious, sub-optimal decision to accommodate short-term goals at the expense of long-term value. The asset is degraded as a result of a modification comprising a conscious non-optimal decision. We take a shortcut, knowing its consequences, and eventually pay its price. 
    \item \textbf{Unintentional}: Sub-optimal decisions based on lack of diligence are unintentional forms of degradation. The asset is degraded as a result of a modification comprising an unconscious non-optimal decision. We are not aware of the other, better alternatives to our decision; therefore, we do not foresee the consequences of selecting an alternative. There is no awareness of the fact that the usability of the asset can be hindered.
    \item \textbf{Entropy}: According to Lehman, ``as a software program is evolved its complexity increases unless work is done to maintain or reduce it'' and ``[it] will be perceived as of declining quality unless rigorously maintained and adapted to a changing operational environment'' \cite{Lehman1979, lehman1996laws}. This ever increasing complexity and perceived quality decline might encompass the degradation of the assets involved in the development of the software system. The degradation that is due to the continuous evolution  of the software, and which is not coming directly from the manipulation of the asset by the developers is entropy. Entropy can introduce degradation in the form of natural decay due to technology and market changes on assets even when we have quality management mechanisms to avoid such degradation.
\end{itemize}   

\begin{figure*}[ht]
\centering
  \includegraphics[width=0.9\textwidth]{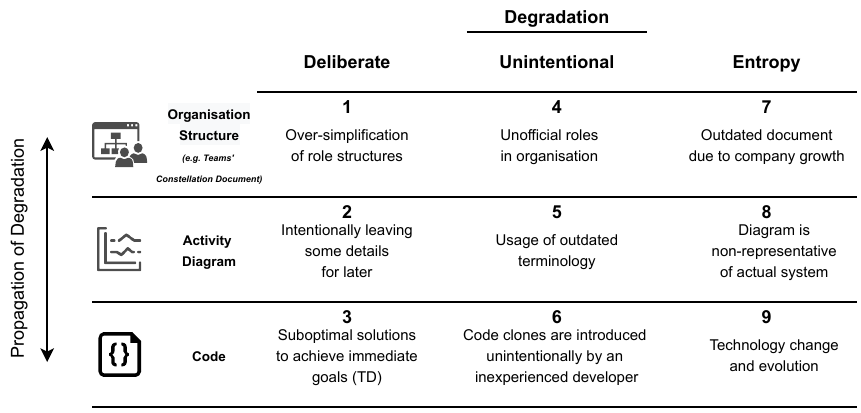}
\caption{There are three sources for asset degradation: deliberate, unintentional, and entropy. The propagation of degradation causes other assets to degrade.}
\label{fig2:degradation}   
\end{figure*}

Here, we provide examples of asset degradation. The numbers in the list correspond to the numbers in Figure~\ref{fig2:degradation}.  

\vspace{2mm}
\noindent \textbf{\textit{Deliberate Degradation:}}

\begin{enumerate}
    \item \textbf{Deliberate Degradation of Organisation Structure}: If the organisation's structure (a team's constellation document) is over-simplified on purpose by associating only the main role to an individual, the asset is deliberately degraded. The saved effort comes at the expense of the structure not representing all access rights and responsibilities, hence inhibiting its use for use cases, where more than the main role of an individual needs to be retrieved.
    \item \textbf{Deliberate Degradation of Activity Diagram}: If the activity diagram is drawn intentionally incomplete to save time (drawn quick-and-dirty), the asset is deliberately degraded.
    \item \textbf{Deliberate Degradation of Code}: If shortcuts are taken to deliver the functionality by incurring TD on code, the asset is deliberately degraded.
\end{enumerate}

\noindent \textbf{\textit{Unintentional Degradation:}}
    
\begin{enumerate}
    \setcounter{enumi}{3}
    \item \textbf{Unintentional Degradation of Organisation Structure}: If the organisation structure, i.e., teams' constellation document, does not include unofficial roles of the employees, the asset is unintentionally degraded.
    \item \textbf{Unintentional Degradation of Activity Diagram}: If outdated, archaic terminology is used to create an activity diagram (i.e., the current state-of-practice terminology is not used), the asset is unintentionally degraded.
    \item \textbf{Unintentional Degradation of Code}: If an inexperienced developer introduces code clones, the asset is unintentionally degraded.
\end{enumerate}
    
\noindent \textbf{\textit{Degradation Due To Entropy:}}

\begin{enumerate}
    \setcounter{enumi}{6}
    \item \textbf{Degradation of Organisation Structure Due To Entropy}: If the organisation structure, i.e., teams' constellation document, is not representative of the current structure of the organisation after the growth of the company, the asset is degraded due to entropy.
    \item \textbf{Degradation of Activity Diagram Due To Entropy}: If the diagram is not representative of the current state of the system after changes have been made and implemented in the system, the asset is degraded due to entropy. 
    \item \textbf{Degradation of Code Due To Entropy}: If the quality of the code is impacted by releases of new versions of third-party libraries, the code can be outdated due to, for example, the usage of deprecated functions and the asset degraded due to entropy, not due to intentional or unintentional design decisions made by developers.
\end{enumerate}

\subsection{Assets' Degradation Can Propagate}

When an asset is degraded, it is likely to influence the value of other dependent assets. We refer to this relation as ``propagation'' of asset degradation. It is important to note that degradation is different from interest. Interest is the term used in the TD metaphor defined as ``the additional cost of the developing new software depending on not-quite-right code ~\cite{Avgeriou2016}.'' Interest is the inflation of the initial cost of repayment (principal) for one asset. At the same time, the propagation of degradation is the impact on other assets caused by a degraded asset.

An example can be the propagation of the degradation of the test base when we deprecate \textit{dead~\footnote{We use here the concept of dead code to refer to versions of functions or services that are never used in the final product or by client applications.}} versions of functions or services~\cite{Sundelin2020}. We can be tempted to deprecate these dead versions instead of removing them since they are never used. The rationale behind that decision is that deprecation will prevent developers from reusing them, provided that nowadays almost every development environment (IDE) will show deprecation information to developers. This might save us the cost of actually removing the code. However, these functions are still tested, and the test cases of the \textit{dead}, deprecated functions or services can be reused as well in other tests, or even the deprecated functions or services can be called directly from integration or system tests to create the mock-up data or the required system state. In those cases, the deprecation information might not be visible for testers since sometimes the integration tests are using XML or other structured formats not supported by IDEs. Testers might continue reusing old, dead, deprecated versions of a function or service for some time, degrading the quality of the test base as a result of a coding design decision.

\subsection{Asset Management}

We define asset management as an umbrella term for the administration of assets and the activities that are related to creating and maintaining them as well as controlling their quality. Asset management considers the explicit control of assets throughout their life-cycle with a particular focus on the assets' degradation and ``emendation'', the conscious removal of degradation.

\subsection{Assets and Software Evolution}

\begin{figure}[ht]
    \centering
      \includegraphics[width=0.35\textwidth]{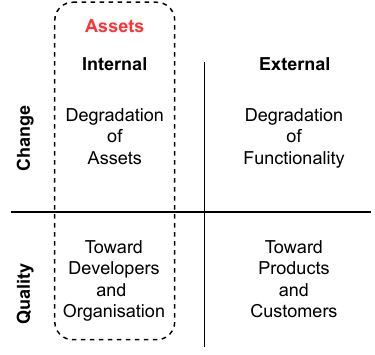}
    \caption{The nature of software evolution: Quality and Change \cite{reussner2019managed}. Assets are within the internal aspect of software evolution.}
    \label{fig:evolution}   
    \end{figure}

Our work is reminiscent of the work of Lehman on software evolution where evolving (E-type) and not-evolving (S-type) systems are differentiated~\cite{Lehman1979,Lehman1980,lehman1996laws}. However, where previous work (e.g.,~\cite{demeyer2008software,reussner2019managed}) focuses on the external aspect of software evolution, and at a system level, our work focuses on the \textit{internal} aspect (where assets are relevant). Figure~\ref{fig:evolution} illustrates the differences of internal and external aspects in respect to the nature of software evolution, i.e., \textit{quality} and \textit{change}~\cite{reussner2019managed}, and highlights how our work differs from already established literature:
 
\begin{itemize}
    \item \textbf{Quality:} While the previous work on the topic mainly considers software evolution on the external quality of the software, i.e. toward the products' quality in use and customers; our work focuses on software evolution on the internal quality of software, toward the developers and the software development organisations.
    \item \textbf{Change:} The four \textit{`Kinds of Software Change'} include \textit{corrective, adaptive, perfective,} and \textit{preventive} modifications~\cite{reussner2019managed}. Such modifications are discussed in the scope of software systems and products and in relation to the degradation of the external functionality. However, our work discusses the degradation of internal assets related to the development process.
\end{itemize}

To the best of our knowledge, despite the existing research on individual, case-based assets, there has not yet been given any proposal for a definition and characterisation of assets and asset degradation in software engineering. 

\section{Future Perspectives for Research and Practice}
A shift in the view on assets, their value, and their management ---starting with a precise and concise terminology--- could enable a new way of considering how the metaphor of debt can be applied in software engineering and, related, building tangible asset taxonomies.

TD, as a metaphor, mapped from economic sciences ---as it is predominantly used today--- runs the risk of not properly reflecting the consequences of sub-optimal decisions. Therefore, it is limited in its application as a concept for capturing the effort necessary to remedy these consequences.

This is mainly due to the fact that the ``interest'' part of TD is often interpreted as covering all consequences of TD. At the same time, it reflects the effort necessary to clear that particular TD in addition to the effort associated with its ``principal''. But since TD is confined to single assets, the complex relationship between various assets is omitted. As defined above, the notion of propagation is tailored to cover exactly this aspect and to better model the inter-dependencies of value-degradation between assets.

Research on traceability focuses on the relationships between certain assets, e.g., between requirements and code, or code and tests. Still, it struggles to provide a holistic view of assets and the relations among them. Rethinking the significance of assets and accordingly adjusting risk mitigation strategies is a first step towards rectifying this misalignment.

In particular, we would like to focus on the following research agenda and practical changes following this work:

\begin{itemize}
    \item Effectively managing the quality of artefacts requires awareness and a profound understanding of which of those artefacts constitute (key) assets. 
    \item As mentioned in Section~\ref{sec:characteristics}, artefacts may qualify as assets without an apparent intention for reuse. This happens when the involved stakeholders fail to predict that an artefact will actually be used more than once. Failing to identify assets early on may cause additional cost, as -- once the artefact has been correctly reassessed to be an asset -- the quality of the asset has not been controlled and it degraded in the meantime. Predicting which eventually actually qualify as assets will be a major future research to reduce additional cost caused by lack of quality control.
    \item The quality of assets can degrade, thus, a depreciation of the value of assets takes place. While this notion of depreciation of software artefacts is not yet prevalent in Software Engineering (as it is in other disciplines, such as economics), it is nevertheless important to manage asset degradation. We hypothesise that there is a close relationship between the degradation of an asset and its depreciation. Asset degradation might increase maintenance costs, and hinder the ability to reuse these assets to create new functionality or services relevant for the users. We can intuitively link these two aspects (i.e., higher maintenance costs and lower reusability) to the depreciation of the degraded assets, however, this is again the subject of further research on the area.
    \item Research on degradation and propagation of degradation is a necessary foundation for controlling assets' quality. 
    \item Once we understand the notion of asset degradation, we can effectively integrate practices for quality control (e.g., in continuous software engineering environments \cite{klotins2022})
\end{itemize}

\section{Conclusion}
In this paper, we provide a definition for assets and asset management, introducing the concept of asset degradation. Based on our experience, after conducting workshops related to the concept of technical debt and asset degradation with several companies, we see that the term \textit{technical debt} can be misleading. We may improve how we view and handle assets, particularly when discussing potential actions in practice, by introducing a coherent set of concepts and the corresponding, concise terminology, such as degradation.

We specifically aspire to construct a taxonomy for assets that provides the means for effective impact analysis to support evidence-driven risk management approaches properly. Those can then take: 
\begin{inparaenum} [i)]
    \item propagation of degradation from one asset to all inter-related assets,
    \item measurement metrics for the degradation on the impacted assets,
    \item evaluation of the severity of the propagated degradation to estimate the true consequences of degradation,
    \item and strategies of emendation for managing the risk and clearing the degradation.
\end{inparaenum}
Our vision is that this work provides a guideline for applying the concept of asset degradation in a practical context and better understanding the complex relationships between the value of assets in practice. At the time of writing this manuscript, we are working on creating a taxonomy of assets.

\section*{Acknowledgements}
We would like to thank Neil Ernst for helpful comments on earlier versions of this manuscript. 